\documentclass[twocolumn,superscriptaddress,aps,preprintnumbers,amsmath,amssymb,nofootinbib,prl]{revtex4}
\usepackage{graphicx}
\usepackage{epstopdf}
\usepackage{dcolumn}
\usepackage{bm}
\usepackage{hyperref}
\def\slashchar#1{\setbox0=\hbox{$#1$} 
\dimen0=\wd0 
\setbox1=\hbox{/} \dimen1=\wd1 
\ifdim\dimen0>\dimen1 
\rlap{\hbox to \dimen0{\hfil/\hfil}} 
#1 
\else 
\rlap{\hbox to \dimen1{\hfil$#1$\hfil}} 
/ 
\fi}

\def\b{\beta}

\def\l{\lambda}

\def\beq{\begin{eqnarray}}
\def\eeq{\end{eqnarray}}

\newcommand{\vev}[1]{ \left\langle {#1} \right\rangle }

\begin{document}
\title{Neutralino Dark Matter in Gauge Mediation in Light of CDMS-II}

\author{Masahiro Ibe}
\affiliation{%
Department of Physics and Astronomy, 
University of California, Irvine, California 92697, USA 
}
\author{Tsutomu T.~Yanagida}
\affiliation{Institute for the Physics and Mathematics of the Universe, University of Tokyo, 
Kashiwa 277-8568, Japan.}



\begin{abstract}
A recent observation of the two candidate events of the dark matter recoiling
at CDMS-II is suggestive of dark matter with a mass not far above 100\,GeV.
We propose a model of  gauge mediated supersymmetry 
breaking where the lightest neutralino is identified as dark matter which
may provide the observed signals.
\end{abstract}

\date{\today}
\maketitle
\preprint{IPMU09-0161}

\subsection*{Introduction}
For more than 70 years, dark matter has eluded direct detection,
and its nature still remains unclear.
However, a recent  observation of the two candidate events of the dark matter recoiling
at CDMS-II may be providing an important hint of the nature of  dark matter\,\cite{CDMSII}.
Especially, the recoil energies detected at $12.3$\,keV and $15.5$\,keV suggest
that the mass of dark matter is not so heavier than $100$\,GeV.

From the theory side in the era of the LHC, the most interesting candidate of dark matter 
is the lightest neutralino in the supersymmetric standard model (SSM).
So far, there have been a lot of works on the direct detection of the neutralino dark matter 
scenario\,\cite{Jungman:1995df}.

From the point of view of supersymmetric model building, however, 
the neutralino dark matter with a mass not so far above 100\,GeV has some tensions.
For example, in gravity mediation,
all the superparticles are expected to have comparable masses.
Such a rather light spectrum, however, predicts too light higgs particles.
Besides, in gravity mediation, it is rather difficult to
suppress the supersymmetric contributions to flavor 
changing neutral current (FCNC) processes.%
\footnote{Both the problems in gravity mediation can be ameliorated  in large cutoff supergravity 
models in Ref.\,\cite{Ibe:2004mp}.
The interpretation of the CDMS-II results in large cutoff supergravity
will be disscussed elsewhere\,\cite{Ibe:2010}.}
On the other hand, in most of the models with gauge mediation,
the gravitino is lighter than the lightest neutralino, and hence,
the lightest neutralino is no more the dark matter candidate,
although the FCNC problem is naturally solved.

In this paper, we propose a model with gauge mediation
where the gravitino is heavier than the lightest neutralino,
while the masses of the sfermions are dominated by
the gauge mediation effects.

\subsection*{Hierarchical gauge mediation}
In order for the lightest neutralino with a mass not so far above $100$\,GeV
to be the lightest superparticle, 
the gravitino mass should be heavier than the lightest neutralino, i.e.,
\begin{eqnarray}
 m_{3/2} = \frac{F}{\sqrt{3} M_{\rm PL}} \gtrsim 100\,{\rm GeV}\ .
\end{eqnarray}
Here, $F$ denotes a supersymmetry breaking vacuum expectation value
and $M_{\rm PL} = 2.4 \times 10^{18}$\,GeV is the reduced Planck scale.
Thus, the supersymmetry breaking expectation value is larger than,
\begin{eqnarray}
\label{eq:susybreaking}
 F\gtrsim 4\times 10^{20}\,{\rm GeV}^2\ .
\end{eqnarray}

On the other hand, the sfermion masses should be dominated 
by the gauge mediated contributions, so that the 
flavor violating masses from gravity mediation are relatively suppressed, i.e.
\begin{eqnarray}
 m_{\rm scalar}^{\rm (GMSB)} \gg m_{3/2} \gtrsim 100\,{\rm GeV}\ .
\end{eqnarray}
In this study, we assume that the gravity mediation contribution
is less than about 1\,\%, i.e.,
\begin{eqnarray}
m_{3/2}^2/m_{\rm scalar}^{\rm (GMSB)2} \lesssim 0.01\ ,
\end{eqnarray}
or equivalently,
\begin{eqnarray}
\label{eq:hierarchical}
 m_{\rm scalar}^{\rm (GMSB)} \gtrsim 1\,{\rm TeV}\ .
\end{eqnarray}

Put it all together, we require a hierarchical spectrum,
\begin{eqnarray}
\{m_{\rm gaugino}^{\rm (GMSB)}, \mu_H  \} 
\lesssim m_{3/2}
\ll m_{\rm scalar}^{\rm (GMSB)}\, , 
\end{eqnarray}
where $\mu_H$ denotes the supersymmetric higgs mixing parameter ($\mu$-term).
Hereafter, we assume that the sfermion masses are not far above $O(1)$\,TeV
to avoid the large hierarchy problem.

Can such a hierarchical spectrum be realized in models with gauge mediation?
In fact, it is generic that the gaugino masses are 
suppressed compared with the sfermion masses in R-symmetric gauge mediation
models\,\cite{Komargodski:2009jf}.

To see the suppression explicitly, let us consider a model of gauge mediation
developed in Refs.\,\cite{Izawa:1997gs,Nomura:1997uu},
where the messenger fields $\psi$, $\bar{\psi}$, $\psi'$ and $\bar{\psi}'$
couple to a supersymmetry breaking chiral superfield, $S = F_S\,\theta^2$, in the superpotential, 
\begin{eqnarray}
\label{eq:GMMODEL}
W = S\psi\bar{\psi} + M_{\rm mess} \psi \bar{\psi}' + M_{\rm mess} \psi' \bar{\psi} + M_{\slashchar R} \psi'\bar{\psi}' \ .
\end{eqnarray}
Here, $M_{\rm mess}$  and $M_{\slashchar R}$ are mass parameters.
The above superpotential possesses the R-symmetry in the limit of vanishing $M_{\slashchar R}$.
In the followings, we assume $F_S = F$ for simplicity, although we can extend our analysis
for more generic cases with $F_S < F$ straightforwardly.

In this model, the sfermion masses are given by,
\begin{eqnarray}
\label{eq:sfermion}
 m_{\rm scalar}^2 \simeq \sum_{a=1,2,3}2C_a\left(\frac{\alpha_a}{4\pi}\right)^2 \left( \frac{F}{M_{\rm mess}} \right)^2\ ,
\end{eqnarray}
where $\alpha_a$ denotes the fine structure constant of the each SSM gauge groups,
and $C_a$ is an order one coefficient which depends on the group representations of the messenger fields.
Here, we have assumed that the mass parameter $M_{\slashchar R}$ is smaller than $M_{\rm mess}$,
which will be our main concern in the following discussion.
From the supersymmetry breaking scale in Eq.\,(\ref{eq:susybreaking}),
the sfermion masses in the TeV range imply
\begin{eqnarray}
\label{eq:mess}
 M_{\rm mess} \simeq 10^{15}\,{\rm GeV}.
\end{eqnarray}

On the other hand, the gaugino masses ($a=1,2,3$) are given by\,\cite{Giudice:1997ni},
\begin{eqnarray}
\label{eq:gaugino}
 m_{a} \simeq b_a \frac{\alpha_a}{4\pi} \times F \times \frac{\partial}{\partial S} \log\left[\det M_\psi\right],
\end{eqnarray}
where $b_a$ denotes an order one coefficient which also depends on the group representations of the messenger fields.
Here, the mass matrix $M_{\psi}$ is defined by,
\begin{eqnarray}
M_{\psi} =
\left(
\begin{array}{cc}
 S   & M_{\rm mess}  \\
M_{\rm mess}   &   M_{\slashchar R}\\
\end{array}
\right)\ ,
\end{eqnarray}
and its determinant is given by,
\begin{eqnarray}
\label{eq:det}
\det M_\psi = S M_{\slashchar R} - M_{\rm mess}^2\ .
\end{eqnarray}
From Eqs.\,(\ref{eq:gaugino}) and (\ref{eq:det}), we immediately find 
that the gaugino mass is vanishing in the limit of $M_{\slashchar R}\to 0$,
even if the supersymmetry breaking chiral field $S$ obtains a spontaneous R-symmetry
breaking scalar expectation value from a supersymmetry breaking sector, i.e. $S = M_R + F_S\,\theta^2$.%
\footnote{
Strictly speaking,  
the gauginos
obtain masses which are suppressed by $|F/M_{\rm mess}^2|^2$ than 
the sfermion masses in Eq.\,(\ref{eq:sfermion})
even in the limit of $M_{\slashchar R} \to 0$. 
However, they are negligibly small for $M_{\rm mess}\simeq 10^{15}$\,GeV.
Those gaugino masses can be important when
the messenger scale is as low as $M_{\rm mess} = O(100)$\,TeV
for $F_S \ll F$.
Even in this case, the gaugino masses are still smaller than the sfermions\,\cite{Ibe:2005xc}, 
and hence, we may explain the hierarchical spectrum without introducing $M_{\slashchar R}$.
A detailed analysis of this case will be given elsewhere.
}
With non-vanishing $M_{\slashchar R}$, the gaugino masses  are given by,
\begin{eqnarray}
m_{a} \simeq b_a \frac{\alpha_a}{4\pi} \times \frac{F }{M_{\rm mess}} \times \frac{M_{\slashchar R}}{M_{\rm mess}}\ .
\end{eqnarray}
Therefore, by choosing an appropriate R-breaking mass $M_{\slashchar R}$, 
we can realize suppressed gaugino masses compared with the sfermion masses.

The suppressed gravitino mass compared with the sfermion masses is also beneficial
to explain the suppressed $\mu$-term.
That is, the $\mu$-term can be generated by the Giudice-Masiero mechanism\,\cite{Giudice:1988yz},
where the $\mu$-term originates from a term in the Kalher potential,%
\footnote{
We may also obtain a similar size of $\mu$-term via a Kahler potential\,\cite{Inoue:1991rk},
\begin{eqnarray}
\label{eq:GMII}
 K = c_H' H_u H_d + h.c.\ ,
\end{eqnarray}
which gives $\mu_H = c_H' m_{3/2}$.
}
\begin{eqnarray}
\label{eq:GM}
 K = \frac{c_H S^{\dagger}}{\sqrt{3}M_{\rm PL}} H_u H_d + h.c.\ .
\end{eqnarray}
Here, $c_H$ is an order one coefficient, and $H_{u,d}$ denote the Higgs doublet superfields.
Then, the resultant $\mu$-term is given by,
\begin{eqnarray}
\mu_H = c_H \times m_{3/2}\ .
\end{eqnarray}
Thus, the suppressed gravitino mass provides the suppressed $\mu$-term.

\subsection*{Discussions}
In the previous section, we have proposed
a model with the lightest neutralino dark matter
which are free from the FCNC problem;
a gauge mediation model with a hierarchical spectrum in Eq.\,(\ref{eq:hierarchical}).
The gravitino mass is in between the lightest neutralino mass and the sfermion masses.
In this model, for example, we may obtain a following parameter set,
\begin{eqnarray}
\label{eq:param}
m_1 &=& 75\,{\rm GeV},\, m_2=350\,{\rm GeV},\,m_3 = 800\,{\rm GeV},\cr
\mu_H &=&125\,{\rm GeV},\, \tan\beta = 10,\, m_{\rm scalar}=2\,{\rm TeV},
\end{eqnarray}
where the parameters are given at the renormalization scale around the electroweak scale.
For simplicity, we have assigned the same masses to all the squarks and sfermions, although
they are not relevant for the direct detection rate below.
With this mass parameter set, the lightest neutralino has a mass $m_{\chi} = 59$\,GeV
and gets sizable higgsino components.
The expected number of events at  direct detection by Ge detectors
is $1.16\times 10^{-2}$/day/kg, ($\sigma_{\chi-p}^{SI} = 2.9\times 10^{-44}$cm$^2$), 
which is consistent with the detection of the two candidate events
in CDMS-II\,\cite{CDMSII}, while the relic density of dark matter 
is consistent with the WMAP observation\,\cite{Komatsu:2008hk}.%
\footnote{In this analysis, we have used {\it micrOMEGAs2.1}\,\cite{Belanger:2008sj}.
The  annihilation process is dominated by an  $s$-channel pole exchange of the light higgs boson 
with a mass of $117$\,GeV.
For a heavier neutralino with $m_{\chi}>m_{Z,W}$,  
the annihilation process is dominated by the modes into $W^+W^-$ and $ZZ$ bosons
which are allowed by the higgsino components of the lightest neutralino.
}

Several comments are in order.
In the above parameter set,  we have not assumed the 
so-called GUT (grand unified theory) relations between the mass parameters.%
\footnote{The gaugino mass not satisfying the GUT relation
is not absolutely necessary to explain the CDMS-II result.
}
In this model, such a spectrum without the GUT relation
can be easily realized in a GUT consist manner.
For example, we may introduce coupling constants between
the messengers and the supersymmetry breaking field which depend
on the vacuum expectation value for spontaneous breaking of the GUT gauge group.
In this case, the coupling constants between  messenger fields and the supersymmetry breaking field
do not necessarily satisfy the GUT relation.
Thus, for example,  the mass ratio $m_1/m_2\simeq 0.21$  in Eq.\,(\ref{eq:param}) 
can be realized when the ratios of the coupling constants of $S Q_M\bar{Q}_M$,
$S \bar{U}_M U_M$ and $S \bar{E}_M E_M$ interaction in Eq.\,(\ref{eq:GMMODEL}) 
are around $1:0.4:0.4$, while keeping the common messenger masses.
Here, $(Q_M,\bar{U}_M,\bar{E}_M)$ and its conjugate representations denote messenger fields
of $\bf10$+$\bf 10*$ representations in terms of the $SU(5)$ GUT gauge group.%
\footnote{As an interesting prediction, the mass ratio $m_1/m_2$ is correlated to the mass ratio between right-handed and left-handed sleptons.
}

As another example, we may also introduce messenger fields which do not consist of the complete representations
of the GUT gauge groups,
such as the messenger fields belonging to the adjoint representations
of $SU(3)$ and $SU(2)$ gauge groups in the SSM which would be in
the adjoint representation of the $SU(5)$ GUT gauge group.%
\footnote{The introduction of the adjoint representations of the $SU(3)$ and $SU(2)$ gauge groups 
does not spoil the coupling unification, though the unification scale is affected\,\cite{Bachas:1995yt}.
As long as the messenger scale is close enough to the so-called GUT scale, $M_{\rm GUT}\simeq 10^{16}$\,GeV,
however, the change of the unification scale is small.
} 
From this messenger sector, there is no contributions to the $m_1$
while $m_{2,3}$ obtain non-vanishing gaugino masses.
Thus, the gaugino masses without the GUT relation can be explained by
introducing multiple messenger fields such as $\bf 5$+$\bf 5^*$ 
representations  in terms of the $SU(5)$ GUT gauge group
with the above adjoint representations\,\cite{Han:1998pa}
(see also Ref.\,\cite{Dudas:2008eq} for a recent related discussion
on the adjoint messengers).%
\footnote{When the mediation scale in Eq.\,(\ref{eq:mess}) is close to the GUT scale,
the gaugino masses also get contributions from the remaining heavier components in the adjoint 
representations of the $SU(5)$ GUT gauge group. 
In this case, we may not need to introduce multiple messenger fields
to explain the relatively small $m_1$ than $m_{2,3}$.
}

Electroweak symmetry breaking may require careful tuning between mass parameters.
In this model, the supersymmetry breaking masses of the Higgs doublets are
in the TeV range at the mediation scale.
For a successful electroweak symmetry breaking,
the one of them should be in a similar size of the $\mu$-term at the electroweak scale.
That is, the supersymmetry breaking  mass at the mediation scale 
should be cancelled by the renormalization group contributions which mainly come from 
the stop mass contributions. 
Furthermore, the Giudice-Masiero mechanism in Eq.\,(\ref{eq:GM}) generates 
the supersymmetry breaking Higgs mixing parameter ($B\mu$-term),
\begin{eqnarray}
  B\mu = c_H m_{3/2}^2\ .
\end{eqnarray}
Thus, the $B\mu$-term is also suppressed than the Higgs soft masses,
which tends to predict rather large $\tan\b$.
A further consistency check of electroweak symmetry breaking will be discussed elsewhere.

The cosmic abundance of the gravitino with a mass in the hundreds GeV range 
is strictly restricted by the constraints on the effects on the Big-Bang-Nucleosynthesis (BBN).
As discussed in Ref.\,\cite{Kawasaki:2008qe}, the reheating temperature after the primordial inflation
is constrained to be lower than $10^{6-7}$\,GeV for $m_{3/2}=O(100)$\,GeV
to suppress the gravitino abundance.
The reheating temperature in this range is too low to explain the baryon asymmetry
via Leptogenesis\,\cite{Fukugita:1986hr}.%
\footnote{The non-thermal Leptogenesis works for a relatively low reheating temperature as low as 
$10^{6-7}$\,GeV\,\cite{Kumekawa:1994gx,Lazarides:1996dv}. }
Therefore, it is a non-trivial question whether there's a consistent scenario of  cosmology with this model
including the baryon asymmetry.

The CP violations of the gaugino masses and the $\mu$- and $B\mu$-terms 
can be unacceptably large when we introduce multiple messengers.
In this case, the phases of the gaugino masses are no more universal,
which cannot be rotated away by using the definitions of the fields.%
\footnote{As briefly mentioned above, we may have a desired hierarchical
spectrum with $M_{\slashchar R}=0$ for $M_{\rm mess} = O(100)$\,GeV.
In such an R-symmetric gauge mediation model,
we can rotate away the phases of the gauginos.
}
Besides, the phases appearing  in $\mu$-term and $B\mu$-term cannot
be rotated away.
Those CP-violating phases lead to, for example, too large electric dipole moment.
A possible way out to this problem is to consider spontaneous CP violation\,\cite{Hiller:2002um},
which eliminate all the CP-phases by symmetry while providing the CKM phase
after spontaneous CP breaking.

Finally, we comment on other possibilities to realize the neutralino dark matter 
in gauge mediation.
As discussed in Refs.\cite{Nomura:2001ub,Shirai:2008qt,Craig:2008vs}, 
the neutralino dark matter is also possible
in gauge mediation when the gravity mediated supersymmetry breaking effects
are suppressed by the so called sequestering mechanisms\,\cite{Randall:1998uk,Luty:2001zv}.%
\footnote{See also Refs.\,\cite{Ibe:2005pj,Ibe:2005qv,Schmaltz:2006qs}
for the later development of the conformal sequestering mechanisms.}
In those attempts, we do not need a hierarchy between the gaugino masses
and the sfermion masses to suppress the flavor violating scalar masses
(see details in the appendix).

\section*{Acknowledgements}
The work  T.T.Y. was supported in
part by World Premier International Research Center Initiative (WPI Initiative), 
MEXT, Japan.

\subsection*{Appendix: Gauge mediation with sequestered gravity mediation}
In this appendix, we give a bare bones outline of a gauge mediated model with sequestered 
gravity mediation which provides the rather light lightest neutralino. 

To be specific, we consider the vector-like supersymmetry breaking model
based on an $SU(2)$ gauge theory with four fundamental representation 
fields $Q_i (i = 1,\cdots, 4)$ and six singlet fields
$Z_{ij} = -Z_{ji} (i, j = 1,\cdots , 4)$\,\cite{Izawa:1996pk,Intriligator:1996pu}.
In this model, the SUSY is dynamically broken when the $Q$'s and $Z$'s couple 
in the superpotential,
\begin{eqnarray}
 W = \l_{ij} Z_{ij}Q_{i}Q_{j}\ , \, (i<j)\ ,
\end{eqnarray}
where $\l_{ij}$ denotes coupling constants.
The supersymmetry is broken as a result of the tension between the $F$-term conditions
of $Z$'s and $Q$'s.

According to Ref.\,\cite{Ibe:2005pj}, we can extend the above supersymmetry
breaking model to a model with conformal sequestering by adding appropriate
number of gauge charged fields with an appropriate gauge group extension.
The extended model flows to the original supersymmetry breaking model
below the mass scale of the newly added charged fields, $M_{\rm seq}$, which is
explicit breaking to the conformal symmetry at the higher energy scale.
Then, the gravity mediation effects are suppressed by,
\begin{eqnarray}
 m_{\rm scalar}^{({\rm gravity}) 2} \sim \left(\frac{M_{\rm seq}}{M_{\rm PL}}\right)^{\beta'}\times m_{3/2}^2\ ,
\end{eqnarray}
when the model is in the vicinity of the infrared fixed point of the extended model around the Planck scale.
Here, $\beta'$ denotes the derivative of the beta function of the gauge coupling constant with respect 
to the fine-structure constant and it is expected to be of the order of one when the model is strongly interacting.
Thus, for example, $M_{\rm seq}\simeq 10^{11}$\,GeV provides enough suppressions to the gravity
mediation effects for $m_{3/2}=O(100)$\,GeV.

Now, let us consider to mediate supersymmetry breaking to the SSM sector.
For that purpose, we again introduce the messenger sector used in the text,
but here, we assume that $M_{\slashchar R}=0$.
Besides, we treat the chiral field $S$ not as a spurious field as in the text,
but as a dynamical field which possesses a cubic term in the superpotential.
Altogether, the messenger sector is given by,
\begin{eqnarray}
\label{eq:GMMODELII}
W = S\psi\bar{\psi} + M_{\rm mess} \psi \bar{\psi}' + M_{\rm mess} \psi' \bar{\psi} + \frac{f}{3} S^3\ ,
\end{eqnarray}
where $f$ denotes a coupling constant.
Furthermore, to connect the messenger sector to the supersymmetry breaking sector,
we use a mechanism in Ref.\,\cite{Nomura:1997ur}
where we gauge a $U(1)$ subgroup of the
global symmetry in the  supersymmetry breaking sector%
\footnote{We impose a global $SU(4)$ symmetry of the supersymmetry breaking sector 
in the limit of the $U(1)$ gauge coupling vanishing.
Then, as pointed out in Ref.\,\cite{Ibe:2005pj}, we have sequestering. 
This sequestering will be maintained for a non-vanishing $U(1)$ gauge coupling as long as the coupling is very small.}
and 
introduce a pair of $U(1)$ gauge charged superfields $E$ and $\bar{E}$.%
\footnote{One may consider to identify $S$ with one of $Z_{ij}$ which is 
responsible for supersymmetry breaking.
However, in the model in Eq.\,(\ref{eq:GMMODELII}), the couplings between $S$ (i.e. $Z$'s) and the messenger
fields are also sequestered and results in too small gaugino mass (see Eq.\,(\ref{eq:spec2})).
}
The supersymmetry breaking effects are transmitted to $S$ via a coupling,
\begin{eqnarray}
 W =k  S E \bar{E}\ ,
\end{eqnarray}
where $k$ is a coupling constant.
As shown in Ref.\,\cite{Nomura:1997ur}, the $S$ field obtains a negative supersymmetry
breaking mass squared, $m_S^2$, at the two-loop level which destabilize the origin of $S$.
As a result, the R-symmetry and the supersymmetry in the messenger sector are broken by,
\begin{eqnarray}
\vev{S} = \frac{m_S}{f}\ ,\quad
\vev{F_S} = \frac{m_S^2}{f}\ .
\end{eqnarray}

The gaugino masses and the sfermion masses are, then, given by,
\begin{eqnarray}
\label{eq:spec2}
 m_{a}\hspace{.3cm}&\simeq& b_a\frac{\alpha_a}{4\pi}\frac{\vev S}{M_{\rm mess}}  \left| \frac{\vev{F_S}}{M_{\rm mess}^2}\right|^2
 \frac{\vev{F_S}}{M_{\rm mess}}\ , 
 \nonumber\\
 m_{\rm scalar}^{2}&\simeq& \sum_a 2 C_a\left(\frac{\alpha_a}{4\pi}\right)^2
 \left(  \frac{\vev{F_S}}{M_{\rm mess}}\right)^2\ .
\end{eqnarray}
Here, we have assumed $\vev S \lesssim M_{\rm mess}$.
Thus, when $\vev{S}$, $M_{\rm mess}$, $\sqrt{F_S}$ are all in the hundreds TeV range,
we can realize 
\begin{eqnarray}
 m_a, \,m_{\rm scalar} = O(100)\,{\rm GeV}-O(1)\,{\rm TeV}\ ,
\end{eqnarray}
although the gaugino masses tend to be suppressed compared to the scalar masses.%
\footnote{The messenger sector without $\psi'$ and $\bar{\psi}'$ fields with a superpotential 
\begin{eqnarray}
W = S\psi\bar{\psi} + \frac{f}{3} S^3\ , 
\end{eqnarray}
works as well.
In this case, the predicted spectrum is no more hierarchical.
}

It should be noted that the scalar masses (especially the slepton masses)
can be smaller than ones in the model discussed in the text,
 since the flavor-violating gravity mediated effects are sequestered away.
Therefore, this model will be preferred 
if we discover relatively light sfermions at the collider experiments such as LHC.

Finally, we mention an advantage of this model.
This model is impervious to the CP-problem,
while the model in the text may suffer from the problem
when we introduce multiple messengers.
That is, in this class of R-symmetric gauge mediation models,
the CP-phases of the coupling constant between the supersymmetry 
breaking field and the messengers are rotated away even in the 
case of the multiple messengers.


\begin{thebibliography}{99}  
\bibitem{CDMSII}
   Z.~Ahmed et.al. [The CDMS Collaboration],
  arXiv:0912.3592 [astro-ph.CO].
\bibitem{Jungman:1995df}
For a review, see  G.~Jungman, M.~Kamionkowski and K.~Griest,
  Phys.\ Rept.\  {\bf 267}, 195 (1996)
  [arXiv:hep-ph/9506380].
\bibitem{Ibe:2004mp}
  M.~Ibe, K.~I.~Izawa and T.~Yanagida,
  Phys.\ Rev.\  D {\bf 71}, 035005 (2005)
  [arXiv:hep-ph/0409203].
\bibitem{Ibe:2010}  
M.~Ibe, K.-I.~Izawa,T.T.~Yanagida, in preparation.
\bibitem{Komargodski:2009jf}
  Z.~Komargodski and D.~Shih,
  JHEP {\bf 0904}, 093 (2009)
  [arXiv:0902.0030 [hep-th]].
\bibitem{Izawa:1997gs}
  K.~I.~Izawa, Y.~Nomura, K.~Tobe and T.~Yanagida,
  Phys.\ Rev.\  D {\bf 56}, 2886 (1997)
  [arXiv:hep-ph/9705228];
\bibitem{Nomura:1997uu}
  Y.~Nomura and K.~Tobe,
  Phys.\ Rev.\  D {\bf 58}, 055002 (1998)
  [arXiv:hep-ph/9708377].
\bibitem{Giudice:1997ni}
  G.~F.~Giudice and R.~Rattazzi,
  Nucl.\ Phys.\  B {\bf 511}, 25 (1998)
  [arXiv:hep-ph/9706540].
\bibitem{Ibe:2005xc}
  M.~Ibe, K.~Tobe and T.~Yanagida,
  Phys.\ Lett.\  B {\bf 615}, 120 (2005)
  [arXiv:hep-ph/0503098].
  
\bibitem{Giudice:1988yz}
  G.~F.~Giudice and A.~Masiero,
  Phys.\ Lett.\  B {\bf 206}, 480 (1988).
\bibitem{Inoue:1991rk}
  K.~Inoue, M.~Kawasaki, M.~Yamaguchi and T.~Yanagida,
  Phys.\ Rev.\  D {\bf 45}, 328 (1992).
\bibitem{Komatsu:2008hk}
  E.~Komatsu {\it et al.}  [WMAP Collaboration],
  Astrophys.\ J.\ Suppl.\  {\bf 180}, 330 (2009)
  [arXiv:0803.0547 [astro-ph]].
\bibitem{Belanger:2008sj}
  G.~Belanger, F.~Boudjema, A.~Pukhov and A.~Semenov,
  Comput.\ Phys.\ Commun.\  {\bf 180}, 747 (2009)
  [arXiv:0803.2360 [hep-ph]].
\bibitem{Bachas:1995yt}
  C.~Bachas, C.~Fabre and T.~Yanagida,
  Phys.\ Lett.\  B {\bf 370}, 49 (1996)
  [arXiv:hep-th/9510094].
\bibitem{Han:1998pa}
  T.~Han, T.~Yanagida and R.~J.~Zhang,
  Phys.\ Rev.\  D {\bf 58}, 095011 (1998)
  [arXiv:hep-ph/9804228].
  \bibitem{Dudas:2008eq}
 E.~Dudas, S.~Lavignac and J.~Parmentier,
 Nucl.\ Phys.\  B {\bf 808} (2009) 237
 [arXiv:0808.0562 [hep-ph]].
\bibitem{Kawasaki:2008qe}
  M.~Kawasaki, K.~Kohri, T.~Moroi and A.~Yotsuyanagi,
  Phys.\ Rev.\  D {\bf 78}, 065011 (2008)
  [arXiv:0804.3745 [hep-ph]].
\bibitem{Fukugita:1986hr}
  M.~Fukugita and T.~Yanagida,
  Phys.\ Lett.\  B {\bf 174}, 45 (1986).
\bibitem{Kumekawa:1994gx}
  K.~Kumekawa, T.~Moroi and T.~Yanagida,
  Prog.\ Theor.\ Phys.\  {\bf 92}, 437 (1994)
  [arXiv:hep-ph/9405337].
\bibitem{Lazarides:1996dv}
  G.~Lazarides, R.~K.~Schaefer and Q.~Shafi,
  Phys.\ Rev.\  D {\bf 56}, 1324 (1997)
  [arXiv:hep-ph/9608256].
\bibitem{Hiller:2002um}
  G.~Hiller and M.~Schmaltz,
  Phys.\ Rev.\  D {\bf 65}, 096009 (2002)
  [arXiv:hep-ph/0201251].

\bibitem{Nomura:2001ub}
  Y.~Nomura and K.~Suzuki,
  Phys.\ Rev.\  D {\bf 68}, 075005 (2003)
  [arXiv:hep-ph/0110040].
\bibitem{Shirai:2008qt}
  S.~Shirai, F.~Takahashi, T.~T.~Yanagida and K.~Yonekura,
  Phys.\ Rev.\  D {\bf 78} (2008) 075003
  [arXiv:0808.0848 [hep-ph]].
\bibitem{Craig:2008vs}
  N.~J.~Craig and D.~R.~Green,
  Phys.\ Rev.\  D {\bf 79}, 065030 (2009)
  [arXiv:0808.1097 [hep-ph]].
\bibitem{Randall:1998uk}
  L.~Randall and R.~Sundrum,
  Nucl.\ Phys.\  B {\bf 557}, 79 (1999)
  [arXiv:hep-th/9810155].
\bibitem{Luty:2001zv}
  M.~Luty and R.~Sundrum,
  Phys.\ Rev.\  D {\bf 67}, 045007 (2003)
  [arXiv:hep-th/0111231].
\bibitem{Ibe:2005pj}
  M.~Ibe, K.~I.~Izawa, Y.~Nakayama, Y.~Shinbara and T.~Yanagida,
  Phys.\ Rev.\  D {\bf 73}, 015004 (2006)
  [arXiv:hep-ph/0506023];
\bibitem{Ibe:2005qv}
  M.~Ibe, K.~I.~Izawa, Y.~Nakayama, Y.~Shinbara and T.~Yanagida,
  Phys.\ Rev.\  D {\bf 73}, 035012 (2006)
  [arXiv:hep-ph/0509229].
\bibitem{Schmaltz:2006qs}
  M.~Schmaltz and R.~Sundrum,
  JHEP {\bf 0611}, 011 (2006)
  [arXiv:hep-th/0608051].
\bibitem{Izawa:1996pk}
  K.~I.~Izawa and T.~Yanagida,
  Prog.\ Theor.\ Phys.\  {\bf 95}, 829 (1996)
  [arXiv:hep-th/9602180];
\bibitem{Intriligator:1996pu}
  K.~A.~Intriligator and S.~D.~Thomas,
  Nucl.\ Phys.\  B {\bf 473}, 121 (1996)
  [arXiv:hep-th/9603158].
\bibitem{Nomura:1997ur}
  Y.~Nomura, K.~Tobe and T.~Yanagida,
  Phys.\ Lett.\  B {\bf 425}, 107 (1998)
  [arXiv:hep-ph/9711220].

  
\end{thebibliography}
\end{document}